\newcommand{\vecj}{\mbox{\boldmath$j$}}
\newcommand{\veck}{\mbox{\boldmath$k$}}
\newcommand{\vecm}{\mbox{\boldmath$m$}}

\newcommand{\vecr}{\mbox{\boldmath$r$}}

\newcommand{\vecv}{\mbox{\boldmath$v$}}

\newcommand{\vecA}{\mbox{\boldmath$A$}}
\newcommand{\vecB}{\mbox{\boldmath$B$}}

\newcommand{\vecD}{\mbox{\boldmath$D$}}
\newcommand{\vecE}{\mbox{\boldmath$E$}}

\newcommand{\vecH}{\mbox{\boldmath$H$}}

\newcommand{\vecex}{\mbox{{\boldmath$\hat{x}$}}}
\newcommand{\vecey}{\mbox{{\boldmath$\hat{y}$}}}
\newcommand{\vecez}{\mbox{{\boldmath$\hat{z}$}}}

\newcommand{\vecer}{\mbox{{\boldmath$\hat{r}$}}}
\newcommand{\vecen}{\mbox{{\boldmath$\hat{n}$}}}
\newcommand{\vecem}{\mbox{{\boldmath$\hat{m}$}}}
\newcommand{\vecephi}{\mbox{{\boldmath$\hat{\varphi}$}}}
\newcommand{\vecetheta}{\mbox{{\boldmath$\hat{\theta}$}}}
\newcommand{\vecerho}{\mbox{{\boldmath$\hat{\rho}$}}}
\newcommand{\vecnl}{\mbox{\boldmath$0$}}

\newcommand{\vecomega}{\mbox{\boldmath$\omega$}}

\newcommand{\dfd}{{\rm d}}

\newcommand{\half}{\frac{1}{2}}

\newcommand{\ie}{{\em i.e.}}
\newcommand{\eg}{{\em e.g.}}
\newcommand{\etal}{{\em et al.}}

\documentclass[12pt]{article}
\usepackage{graphicx}
\begin{document}

\title{New theorem of classical electromagnetism: equilibrium magnetic field and current density are zero inside ideal conductors}

\author{Miguel C. N. Fiolhais\thanks{LIP-Coimbra, Department of
Physics, University of Coimbra, 3004-516 Coimbra, Portugal} \\   \and Hanno Ess\'en\thanks{Corresponding author, e-mail: hanno@mech.kth.se, Department of Mechanics, KTH, 100 44 Stockholm, Sweden} \\
   \and  C. Provid\^encia\thanks{Centre for Computational Physics, Department of
Physics, University of Coimbra, 3004-516 Coimbra, Portugal} \\
\and Arne B. Nordmark\thanks{Department of Mechanics, KTH, 100 44 Stockholm, Sweden} }

\maketitle

\begin{abstract}
We prove a theorem on the magnetic energy minimum in a system of perfect, or ideal, conductors. It is analogous to Thomson's theorem on the equilibrium electric field and charge distribution in a system of conductors. We first prove Thomson's theorem using a variational principle. Our new theorem is then derived by similar methods. We find that magnetic energy is minimized when the current distribution is a surface current density with zero interior magnetic field; perfect conductors are perfectly diamagnetic. The results agree with currents in superconductors being confined near the surface. The theorem implies a generalized force that expels current and magnetic field from the interior of a conductor that loses its resistivity. Examples of solutions that obey the theorem are presented.
\end{abstract}

\section{Introduction}
Thomson's theorem states that electric charge density on a set of fixed conductors at static equilibrium is distributed on the surface of the conductors in such a way that the interior electric field is zero and the surface electric field is normal to the surface. Here we will prove an analogous theorem on the magnetic field and current distribution in ideal conductors. We find that a stationary current density must distribute itself on the surface of the conductors in such a way that the interior magnetic field is zero while the surface magnetic field is perpendicular to both the current density and the surface normal.

When W.\ Thomson (Lord Kelvin) derived his theorem in 1848 a magnetic analog did not seem interesting since conductors with zero resistivity were unknown. Since the discovery of superconductivity in 1911 this has changed and stationary current distributions in type I superconductors, below the critical field, indeed obey our theorem. In spite of this the theorem is not stated, hardly even hinted at, in the literature. We wish to emphasize that even though the theorem applies to superconductors, phase transitions, statistical mechanics, or thermodynamics are irrelevant. Zero resistivity is simply assumed, not explained or derived. The theorem is purely a consequence of classical electromagnetism.

The outline of this article is as follows. We start by deriving Thomson's theorem using a variational principle. We then derive our minimum magnetic energy theorem in an analogous way and discuss previous work on the problem. After that an illuminating example is presented in which the energy reduction due the interior field expulsion can be calculated explicitly. An Appendix gives further motivation and explicit solutions illustrating the theorem for simple systems.

\section{Energy minimum theorems}
Electromagnetic energy can be written in a number of different
ways. Here we will assume that there are no microscopic dipoles so
that distinguishing between the $\vecD, \vecH$ and $\vecE, \vecB$
fields is unnecessary. That this is valid when treating the
Meissner  effect in type I superconductors is stressed by Carr
\cite{carr}. Relevant energy expressions are then,
\begin{eqnarray}\label{eq.em.energy.field.form}
E_{\rm e} + E_{\rm m} &=& \frac{1}{8\pi}\!\int (\vecE^2 + \vecB^2)\dfd V \\
\label{eq.em.energy.source.pot.form} &=& \half\int
\left(\varrho\phi +
\frac{1}{c}\vecj\cdot\vecA\right)\dfd V  = \\
\label{eq.em.energy.source.only.form}   &=& \half\int\!\!\int
\left(\frac{\varrho(\vecr)\varrho(\vecr') +
\frac{1}{c^2}\vecj(\vecr)\cdot\vecj(\vecr')}{|\vecr-\vecr'|}
\right)\dfd V\dfd V' .
\end{eqnarray}
Here the first form is always valid while the two following
assume quasi statics, \ie\ essentially negligible radiation.

According to Thomson's theorem, the electric charge on a set of conductors distributes itself on the conductor surfaces thereby minimizing the electrostatic energy. W.\ Thomson did not present a formal mathematical proof but such proofs may be found in most classic textbooks \cite{BKjackson3,BKcoulson,BKpanofsky,BKlandau8}. A recent derivation of the theorem in its differential form is by Bakhoum \cite{bakhoum}. A derivation based on a variational principle can be found in the textbook by Kovetz \cite{BKkovetz2}. A different approach also based on a variational principle, is presented below. Thomson's result is widely known and is useful in many applications. It has \eg\ been used to determine the induced surface charge density \cite{sancho&al,donolato}, and in the tracing and the visualization of curvilinear squares field maps \cite{bakhoum}. Other applications range from interesting teaching tools \cite{brito} to useful computational methods such as Monte Carlo energy minimization \cite{sancho}.

There are various similarities between electrostatics and magnetostatics, or quasi-statics,
but for resistive media the magnetic field due to current dissipates\footnote{Note that we are {\em not} concerned with magnetism due to microscopic dipole density.}. For {\em perfect}, or {\em ideal} conductors, however, there should be something corresponding to the magnetic version of Thomson's theorem. Indeed, below we will prove a theorem analogous to that of Thomson: \emph{Magnetic energy is minimized by surface current distributions such that the magnetic field is zero inside while the surface field is normal to the current and the surface normal.} Energy conservation is assumed restricting the validity to perfect, or ideal, conductors. Previously somewhat similar results have appeared in the literature \cite{karlsson,badiamajos} and we discuss those below.

\subsection{Thomson's theorem}
In equilibrium, the electrostatic energy functional for a system
of conductors surrounded by vacuum, may be written as\footnote{The
ultimate motivation for this specific form of the energy
functional, and the corresponding one in the magnetic case, is
that they lead to simple final equations.},
\begin{equation}\label{eq.elstat.energy.miguel.form}
E_{\rm e} = \int_{V} \left[  \varrho \phi - {1 \over 8\pi} \left
(\nabla \phi \right ) ^2 \right ] \dfd V ,
\end{equation}
by combining the electric parts of (\ref{eq.em.energy.field.form})
and (\ref{eq.em.energy.source.pot.form})  and using
$\vecE=-\nabla\phi$. We now split the integration region into the
volume of the conductors, $V_{\rm{in}}$, the exterior volume,
$V_{\rm{out}}$, and the boundary surfaces $S$,
\begin{equation}\label{eq.Ee.split.vol}
E_{\rm e}  =  \int_{V_{\rm{in}}} \left[ \varrho \phi - {1 \over
8\pi} \left (\nabla \phi \right )^2 \right ] \dfd V
-\int_{V_{\rm{out}}} {1 \over 8\pi} \left (\nabla \phi \right )^2
\dfd V  + \int_{S} \sigma \phi \,\dfd S ,
\end{equation}
where $\sigma$ is the surface charge distribution.

We now use, $(\nabla \phi )^2 = \nabla \cdot (\phi \nabla \phi)-
\phi \nabla^2 \phi$, and rewrite  the divergencies using Gauss
theorem. The energy functional then becomes,
\begin{eqnarray} E_{\rm e} = \int_{V_{\rm{in}}} \left [ \varrho \phi +
{1 \over 8\pi}  \phi\, \nabla^2 \phi  \right ] \dfd V
    -\int_{V_{\rm{out}}} {1 \over 8\pi} \phi\, \nabla^2 \phi \,
\dfd V \nonumber\\
+\int_{S} \left [ \sigma \phi - {1 \over 8\pi} \phi\, \vecen
\cdot \left( \nabla^+ \phi - \nabla^- \phi \right )\right ] \,\dfd
S ,
\end{eqnarray}
where $\nabla^+$ and $\nabla^-$ are the gradient operators at the
surface  in the outer and inner limits, respectively. The
total charge in each conductor is constant and restricted to the
conductor volume and surface. We handle this constraint by
introducing a Lagrange multiplier $\lambda$.
Infinitesimal variation of the energy then gives,
\begin{eqnarray}
\delta E_{\rm e} &=&  \int_{V_{\rm{in}}} \left [ \delta \phi
\left( \varrho + \frac{1}{4\pi} \nabla^2 \phi \right) + \delta
\varrho \left ( \phi - \lambda \right ) \right ] \dfd V
    +\int_{V_{\rm{out}}} \delta \phi\, \frac{1}{4\pi}
\nabla^2 \phi\; \dfd V \nonumber \\
&&+ \int_{S} \left \{ \delta \phi \left[ \sigma + \frac{1}{4\pi}
\vecen \cdot \left( \nabla^+ \phi - \nabla^- \phi \right )
\right] + \delta \sigma \left( \phi - \lambda \right) \right \}
\,\dfd S.
\end{eqnarray}
From this energy minimization, the Euler-Lagrange equations
become:
\begin{equation}
\label{eq.interior.pot}
V_{\rm{in}}:\;\; \left\{
\begin{array}{l}
\nabla^2 \phi = -4\pi \varrho \\
\phi = \lambda \\
\end{array} \right.
\label{potencial}
\end{equation}
\begin{equation}
S: \,\,\,\,\left\{
\begin{array}{l}
-\vecen \cdot \left( \nabla^+ \phi - \nabla^- \phi \right ) = 4\pi \sigma\\
\phi = \lambda \\
\end{array} \right.
\label{surface}
\end{equation}
\begin{equation}
V_{\rm{out}}:\hskip 0.5cm \nabla^2 \phi = 0 \label{laplace}
\end{equation}
According to equations (\ref{eq.interior.pot}) the potential is constant inside the
conductor in the minimum energy state, and therefore $\varrho=0$ there. Equations (\ref{surface}) mean the electric
charge is distributed on the surface in such a way that the potential is constant there. The second of eqs.\ (\ref{eq.interior.pot})
implies that $\nabla^- \phi=\vecnl$ and the first of eqs.\ (\ref{surface}) then implies that $\vecen\cdot\vecE_+ = 4\pi \sigma$. This concludes the proof of Thomson's theorem.

\subsection{Minimum magnetic energy theorem}\label{sec.min.mag.energy}
A similar procedure will now be applied to the magnetic field. We
write the magnetic energy functional for a time independent
magnetic field as,
\begin{equation}\label{eq.mag.energy.functional}
E_{\rm m} =  \int_{V} \left [\frac{1}{c} \vecj \cdot \vecA - {1
\over 8\pi} \left (\nabla \times \vecA \right ) ^2  \right ]\dfd V
,
\end{equation}
\ie\ as two times the form (\ref{eq.em.energy.source.pot.form}) of
the magnetic energy minus the form
(\ref{eq.em.energy.field.form}), using $\vecB=\nabla\times\vecA$.
As before we split the volume into the volume interior to
conductors, the exterior vacuum, and the surface at the
interfaces, and write,
\begin{eqnarray} \nonumber
E_{\rm m}  &=& \int_{V_{\rm{in}}} \left [ \frac{1}{c} \vecj \cdot
\vecA - {1 \over 8\pi} \left (\nabla \times \vecA \right ) ^2
\right ] \dfd V \\ & & - \int_{V_{\rm{out}}} {1 \over 8\pi} \left
(\nabla \times \vecA \right ) ^2 \dfd V +\frac{1}{c} \int_{S}
\veck \cdot \vecA \,\dfd S ,
\end{eqnarray}
where $\veck$ is the surface current density. We now use the
identity,
\begin{equation}\label{eq.vector.identity.A}
\left (\nabla \times \vecA \right ) ^2 = \nabla \cdot \left [\vecA
\times \left(\nabla \times \vecA \right )\right ] +\vecA \cdot
\left [\nabla \times \left(\nabla \times \vecA \right ) \right ] ,
\end{equation}
and then use Gauss theorem to rewrite the divergence terms. The
energy functional then becomes:
\begin{eqnarray}
E_{\rm m} &=&  \int_{V_{\rm{in}}} \left \{ \frac{1}{c} \vecj \cdot
\vecA - {1 \over 8\pi}  \vecA \cdot \left [\nabla \times
\left(\nabla \times \vecA \right ) \right ]  \right \} \dfd V
\nonumber \\ \label{variational} & & -\int_{V_{\rm{out}}} {1 \over
8\pi}  \vecA \cdot \left [\nabla
\times \left(\nabla \times \vecA \right ) \right ] \dfd V  \\
\nonumber &&+ \int_{S} \left \{ \frac{1}{c} \veck \cdot \vecA - {1
\over 8\pi} \vecA \cdot \left[ \vecen \times \left( \nabla^+
\times \vecA - \nabla^- \times \vecA \right) \right] \right \}
\,\dfd S .
\end{eqnarray}
As in the electric case, constraints must be imposed. Due to charge conservation, the electric current density must obey the continuity equation, for zero charge density, both inside and on the surface of the conductors \cite{mcallister}:
\begin{eqnarray}
 \nabla \cdot \vecj &=& 0  \\
 \nabla_S \cdot \veck &=& 0
\end{eqnarray}
where $\nabla_S$ is the surface gradient operator. Notice that these
constraints are local, not global. In other words, the relevant Lagrange
multiplier is not constant but a scalar field $\lambda(\vecr)$.
Using this, infinitesimal variation of the magnetic energy gives,
\begin{eqnarray} \nonumber
&& c\, \delta E_{\rm m} =  \int_{V_{\rm{in}}} \left \{ \delta
\vecA \cdot \left(  \vecj - \frac{c}{4\pi}\left [\nabla \times
\left(\nabla \times \vecA \right ) \right ] \right) + \delta \vecj
\cdot \left( \vecA-\nabla\lambda \right) \right \} \dfd V  \\
 &&-\int_{V_{\rm{out}}} \delta \vecA \cdot
\frac{c}{4\pi}\left [\nabla \times \left(\nabla \times \vecA
\right ) \right ] \dfd V  \\&& + \int_{S} \left\{ \delta \vecA
\cdot \left[  \veck + \frac{c}{4\pi} \vecen \times \left(
\nabla^+ \times \vecA - \nabla^- \times \vecA \right ) \right] +
 \delta \veck \cdot \left( \vecA - \nabla_S \lambda
\right) \right \} \,\dfd S . \nonumber
\end{eqnarray}
Equating this to zero we find that,
\begin{equation}
V_{\rm{in}}:\,\, \left\{
\begin{array}{l}
\nabla \times \left(\nabla\times \vecA\right)=\nabla \times
\vecB= \frac{4\pi}{c} \vecj \; , \\
\vecA = \nabla\lambda  \; , \\
\end{array} \right.
\label{potencial2}
\end{equation}
and,
\begin{equation}
S: \,\,\,\,\left\{
\begin{array}{l}
\veck =   \frac{c}{4\pi} \vecen \times
\left( \nabla^+ \times \vecA - \nabla^- \times \vecA \right ) \; , \\
\vecA = \nabla_S\lambda \; ,\\
\end{array} \right.
\label{surface2}
\end{equation}
and,
\begin{equation}
V_{\rm{out}}:\;\;\;  \nabla \times \left(\nabla\times
\vecA\right)=\nabla \times \vecB = 0 , \label{laplace_mag}
\end{equation}
are the Euler-Lagrange equations for this energy functional.

According to eq.\ (\ref{potencial2})
$\vecB=\nabla\times\nabla\lambda=\vecnl$, so the magnetic field
must be zero in $V_{\rm{in}}$. Consequently also the volume
current density is zero, $\vecj=\vecnl$, inside the conductor, in
the minimum energy state.
\begin{figure}[ht]
\centering
\includegraphics[width=200pt]{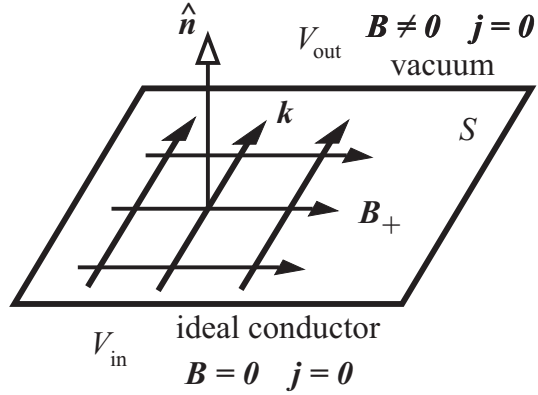}
\caption{\small Some of the results of our theorem on the current density and magnetic field of an ideal conductor at minimum energy are illustrated here. Here $\vecB_+$ is the magnetic field on the outside of the surface $S$ with surface unit normal $\vecen$. The bulk current density $\vecj$ is zero, only the surface current density $\veck$ is non-zero.
\label{fig.MinMagTheorem}}
\end{figure}

\subsubsection{Surface currents} Now consider the results for the
surface, eq.\ (\ref{surface2}). Our results from $V_{\rm{in}}$
show that $\nabla^- \times \vecA=\vecnl$, so the equation reads,
\begin{equation}\label{eq.on.surface}
\veck =   \frac{c}{4\pi} \vecen \times \left( \nabla^+ \times
\nabla_S\lambda  \right ).
\end{equation}
Let us introduce a local Cartesian coordinate system with origin
on the surface, such that the surface is spanned by $\vecex,
\vecey$ with unit normal $\vecen=\vecez=\vecex\times\vecey$.
Assuming that the surface is approximately flat we then have that,
\begin{equation}\label{eq.grad.operators}
\nabla_S = \vecex \frac{\partial}{\partial x} +
\vecey\frac{\partial}{\partial y},\;\;\mbox{and,}\;\; \nabla^+
=\vecex\frac{\partial}{\partial x} +
\vecey\frac{\partial}{\partial y} +
\vecen\frac{\partial}{\partial z^+} =\nabla_S
+\vecen\frac{\partial}{\partial z^+}.
\end{equation}
Since, $\vecA = \nabla_S \lambda$, the vector potential is tangent
to the conducting surface and we get,
\begin{eqnarray}
& \displaystyle \frac{4\pi}{c}\veck = \vecen\times \left[
\left(\nabla_S +\vecen\frac{\partial}{\partial z^+} \right)
\times \nabla_S \lambda(x,y,z)  \right] &  \\ & \displaystyle =
\vecen\times \left[  \vecen\frac{\partial}{\partial z^+} \times
\left( \vecex\frac{\partial\lambda}{\partial x} +
\vecey\frac{\partial\lambda}{\partial y} \right)  \right]  =
\vecen\times \left(  \vecen\frac{\partial}{\partial z^+}
 \times \vecA  \right), &
\end{eqnarray}
for the surface current density. Rewriting the triple vector
product we find,
\begin{equation}\label{eq.triple.prod.k}
\frac{4\pi}{c}\veck = \frac{\partial}{\partial z^+}
\left[(\vecen\cdot\vecA)\vecen -
(\vecen\cdot\vecen)\vecA\right] = -\frac{\partial\vecA}{\partial
z^+} ,
\end{equation}
so the surface current density is parallel to the outside normal
derivative of the vector potential. We note that this agrees with
the well known result \cite{BKlandau8},
\begin{equation}\label{eq.surf.curr.mag.field}
\frac{4\pi}{c}\veck = \vecen \times (\vecB_+ - \vecB_-),
\end{equation}
for the case of zero interior field ($\vecB_- =\vecnl$). Our results are summarized in Fig.\ \ref{fig.MinMagTheorem}.

\subsubsection{External fields}
Neither Thomson's theorem nor our minimum magnetic energy theorem are formally valid for conductors in constant external fields. In both cases, however, such a situation can be regarded as a limiting case. In the case of Thomson's theorem one can include two large, distant, and oppositely charged parallel conducting plates. A small system of conductors between these can then approximately be regarded as in an electric field that approaches a constant external field at large distance. In a similar way the set of perfect conductors can be thought of as inside two large perfectly conducting Helmholtz coils (tori) which provide an approximately constant external magnetic field at large distance.

\subsection{Previous work}
The fact that there is an energy minimum theorem for the magnetic
energy of ideal, or perfect, conductors, analogous to Thomson's
theorem, is not entirely new. In an interesting, but difficult and
ignored, article by Karlsson \cite{karlsson} such a theorem is
stated. Karlsson, however, restricts his theorem to conductors
with holes in them. In the electrostatic case charge conservation
prevents the energy minimum from being the trivial zero field
solution. In our magnetic ideal conductor case the corresponding
conservation law is the conservation of magnetic flux through a
hole \cite{dolecek&delaunay,hehl05}. As long as {\it one} conductor of the system has a hole with conserved flux there will be a non-trivial magnetic field. To
require that all conductors of the system have holes, as Karlsson
does, therefore seems unnecessarily restrictive. One of Karlsson's
results is that a the current distribution on a superconducting
torus minimizes the magnetic energy.

A result by Bad{\'i}a-Maj{\'o}s \cite{badiamajos} comes even
closer to our own and we outline it here. The current density is
assumed to be of the form,
\begin{equation}\label{eq.curr.dens.parts}
\vecj = q n \vecv ,
\end{equation}
where $q$ is the charge of the charge carriers and $n$ is their
number density. The time derivative is then given by,
\begin{equation}\label{eq.time.deriv.j}
\frac{\dfd \vecj}{\dfd\, t} =\frac{qn}{m}\, m \frac{\dfd
\vecv}{\dfd\, t} =\frac{qn}{m} \left(q\vecE +
\frac{q}{c}\vecv\times\vecB \right) = \frac{q^2 n}{m} \vecE +
\frac{q}{mc} \vecj\times\vecB ,
\end{equation}
assuming that only the Lorentz force acts (ideal conductor). We
now recall Poynting's theorem \cite{BKlandau2} for the time
derivative of the field energy density of a system of charged
particles,
\begin{equation}\label{eq.poyntings.theorem}
\frac{\dfd }{\dfd\, t} \left(\frac{\vecE^2 + \vecB^2}{8\pi}
\right) = -\vecj\cdot\vecE -\nabla\cdot\left(\frac{c}{4\pi}
\vecE\times\vecB\right).
\end{equation}
The first term on the right hand side normally represents
resistive energy loss. Here we use the result for $\vecE$ from
eq.\ (\ref{eq.time.deriv.j}),
\begin{equation}\label{eq.E.from.j.and.j.dot}
\vecE = \frac{m}{q^2 n} \frac{\dfd \vecj}{\dfd\, t}
-\frac{1}{qnc}\vecj\times\vecB,
\end{equation}
and get,
\begin{equation}\label{eq.E.dot.j}
\vecj\cdot\vecE =\frac{m}{q^2 n}\, \frac{\dfd \vecj}{\dfd\,
t}\cdot\vecj = \frac{\dfd}{\dfd\, t} \left(\frac{m}{2 q^2 n}
\vecj^2 \right).
\end{equation}
This is thus the natural form for this term for perfect
conductors. We insert it into (\ref{eq.poyntings.theorem}),
neglect radiation, and assume that $\vecE^2 \ll \vecB^2$. This
gives us,
\begin{equation}\label{eq.time.deriv.const.quant}
\frac{\dfd}{\dfd\, t} \left(\frac{1}{8\pi}\vecB^2 + \frac{m}{2 q^2
n} \vecj^2 \right) = 0.
\end{equation}
Finally inserting, $\vecj =(c/4\pi) \nabla\times\vecB$, here,
gives,
\begin{equation}\label{eq.energy.for.perf.cond.B}
E_B =\frac{1}{8\pi}\left[ \int_V \vecB^2 \dfd V + \int_{V_{\rm{in}}}
\frac{mc^2}{4\pi q^2 n} (\nabla\times\vecB)^2 \dfd V \right],
\end{equation}
for the conserved energy, after integration over space and time.

Bad{\'i}a-Maj{\'o}s \cite{badiamajos} then notes that this energy
functional implies flux expulsion from superconductors. Variation
of the functional gives the London equation \cite{london&london},
\begin{equation}\label{eq.london.eq}
\vecB + \frac{1}{4\pi}\frac{mc^2}{q^2 n}
\nabla\times(\nabla\times\vecB) = \vecnl.
\end{equation}
Bad{\'i}a-Maj{\'o}s, chooses not to point out that this classical derivation of flux expulsion is in conflict with frequent text book statements to the effect that no such classical result exists. Further work by Bad{\'i}a-Maj{\'o}s \etal\ \cite{badiamajos&al}  on variational principles for electromagnetism in conducting materials should be noted.

Finally we should mention the pioneering work by Woltjer \cite{woltjer} on energy extremizing properties of, so called, force free magnetic fields. In plasma physics a lot of further work has been done in that tradition. It has, however, not been concerned with currents and fields inside or on the boundary of bounded domains separated by vacuum, as we are here.

\section{Ideally conducting sphere in external field}\label{sec.sphere.in.sphere}
We now know that magnetic energy minimum occurs when current flows only on the surface and the magnetic field is zero inside. Let us consider a perfectly conducting sphere surrounded by a fixed constant external magnetic field. Here we will calculate the surface current needed to exclude the magnetic field from the interior and how much the total magnetic energy $E_m$ is then reduced. In order to exclude a constant external field $\vecB_e$ from its interior the currents on the sphere must obviously produce an interior magnetic field $\vecB_i = -\vecB_e$, thereby making the total field $\vecB = \vecB_e + \vecB_i$ zero in the interior. We will use that a constant field is produced inside a sphere by a current distribution due to rigid rotation of a constant surface charge density \cite{BKgriffiths}.

\subsection{Energy of the external field}\label{subsec.const.ext.field.sphere}
For the total magnetic field to have a finite energy $E_m$ we can not
assume that the constant external field extends to infinity.
Instead of using Helmholtz coils to produce it we simplify the mathematics and
produce our external field by a spherical shell of current that is
equivalent to a rigidly rotating current distribution on the
surface. This can be done in practice by having as set of rings
representing closely spaced longitudes on a globe with the right
amount of current maintained in each of them. Such a spherical
shell of rigidly rotating charge produces a magnetic field that is
constant inside the sphere and a pure dipole field outside the
sphere (see Fig.\ \ref{FieldLinesRotSphere}):
\begin{figure}[ht]
\centering
\includegraphics[width=150pt]{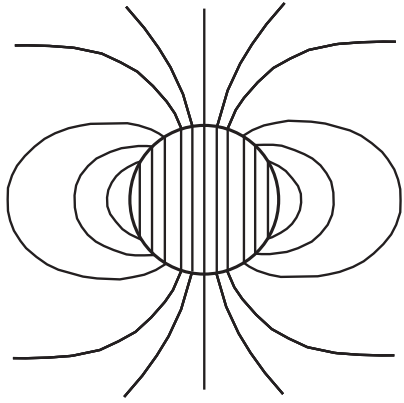}
\caption{\small The field lines of the field of eq.\
(\ref{eq.mag.field.rot.sphere.shell})  for a nonmagnetic sphere
with a rigidly rotating homogeneous surface charge density.
\label{FieldLinesRotSphere}}
\end{figure}
\begin{equation}\label{eq.mag.field.rot.sphere.shell}
\vecB_e(\vecr) = \left\{ \begin{array}{ll} \displaystyle \frac{ 2
 \vecm }{R^3}\, & \mbox{for $r\le R$} \\
   &  \mbox{  } \\
\displaystyle \frac{ 3 (\vecm\cdot\vecer) \vecer - \vecm
}{r^3}\, & \mbox{for $r>R$}
\end{array} \right.
\end{equation}
Here $r=|\vecr|$ and the center of the sphere is at the origin. If
$Q$ is the total rotating surface charge and $\vecomega$ its
angular velocity,
\begin{equation}\label{eq.dipole.vec.of.rot.charge}
\vecm = \frac{Q R^2}{3c} \vecomega ,
\end{equation}
see eq.\ (\ref{eq.middle.field}) below. It is now easy to
calculate the magnetic energy of this field. One finds\footnote{If
(\ref{eq.dipole.vec.of.rot.charge}) is inserted for $\vecm$ here
we get $E_m^0=(\omega/c)^2 R Q^2/9$ which is equal to $E_m$ of eq.\
(\ref{eq.energy.thick.rot.shell}) below, for $\xi=1$,
corresponding to surface current only, as it should.},
\begin{equation}\label{eq.energy.mag.field.rot.sphere}
E_m^0  = \frac{1}{8\pi}\left( \int_{r<R} \vecB_e^2 \, \dfd V + \int_{r>R}
\vecB_e^2 \, \dfd V
\right)=\left(\frac{2}{3}+\frac{1}{3}\right)\frac{\vecm^2}{R^3} =
\frac{m^2}{R^3}.
\end{equation}
Inside this sphere, which is assumed to maintain a constant
current density on its surface, we now place a smaller perfectly
conducting sphere.

\subsection{Magnetic energy of the two sphere system}\label{sec.two.sphere.syst}
We assume that the small sphere in the middle of the big one has
radius $a<R$ and that it also produces a magnetic field by a
rigidly rotating charged shell on its surface. We denote its
dipole moment by $\vecm_i$ so that its total energy would be,
\begin{equation}\label{eq.energy.inner.alone}
E_m^i = \frac{\vecm_i^2}{a^3} =  \frac{m_i^2}{a^3},
\end{equation}
if it was far from all other fields, according to our previous result
(\ref{eq.energy.mag.field.rot.sphere}). We now place the small
sphere inside the large one and assume that $\vecm_i$ makes an
angle $\alpha$ with $\vecm=m\, \vecem$,
\begin{equation}\label{eq.scal.prod.dip.vecs}
\vecm\cdot\vecm_i = m\, m_i \cos\alpha.
\end{equation}
The total energy of the system is now,
\begin{equation}\label{eq.tot.energy.gen.exp}
E_m =\frac{1}{8\pi}\int (\vecB_e + \vecB_i)^2 \dfd V = E_m^0 + E_m^i + E_c
,
\end{equation}
where the coupling (interaction) energy is,
\begin{equation}\label{eq.interact.energy.def}
 E_c =\frac{1}{4\pi}\int \vecB_e \cdot\vecB_i\; \dfd V .
\end{equation}
This integral must be split into the three radial regions: $0
\le r < a$, $a \le r < R$, and $R \le r$. The calculations are
elementary using spherical coordinates. The contribution from the
inner region is,
\begin{equation}\label{eq.interact.energy.1}
 E_{c1} =\frac{1}{4\pi}\int_{r<a} \vecB_e \cdot\vecB_i\, \dfd V =
 \frac{4}{3}\frac{m m_i}{R^3} \cos\alpha .
\end{equation}
The middle region, where there is a superposition of a dipole field from the small
sphere and a constant field from the big one, contributes zero:
$E_{c2}=0$. The outer region gives $E_{c3} =(2/3) m m_i
\cos\alpha/R^3$. Summing up one finds,
\begin{equation}\label{eq.interact.energy.expl}
E_c = 2 \frac{m m_i}{R^3} \cos\alpha ,
\end{equation}
for the magnetic interaction energy of the two spheres.

\subsection{Minimizing the total magnetic energy}
The total magnetic energy of the system discussed above is thus,
\begin{equation}\label{eq.tot.energy.as.func}
E_m(m_i,\alpha) = E_m^0+E_m^i+E_c =  \frac{m^2}{R^3} + \frac{m_i^2}{a^3}
+ 2 \frac{m m_i}{R^3} \cos\alpha .
\end{equation}
We assume that $m$ and $m_i$ are positive quantities. This means
that as a function of $\alpha$ this quantity is guarantied to have
its minimum when $\cos\alpha = -1$, \ie\ for $\alpha = \pi$. Thus,
at minimum, the dipole of the inner sphere has the opposite
direction to that of the constant external field, $\vecm_i = -m_i\vecem$.

Now assuming $\alpha = \pi$ we can look for the minimum as a
function of $m_i$. Elementary algebra shows that this minimum is
attained for,
\begin{equation}\label{eq.min.ma}
m_i =\left( \frac{a}{R} \right)^3 m\, \equiv m_{i\,\rm min}.
\end{equation}
The magnetic field in the interior of the inner sphere ($r<a$) is then,
\begin{equation}\label{eq.mag.field.interior.min}
\vecB_e + \vecB_i = \frac{ 2\vecm }{R^3} + \frac{ 2\vecm_{i\,\rm min} }{ a^3 }
= \left( \frac{2 m }{ R^3 } - \frac{ 2 m_{i\,\rm min}  }{ a^3 }
\right) \vecem
  = \vecnl ,
\end{equation}
so it has been {\it expelled}. The minimized energy
(\ref{eq.tot.energy.as.func}) of the system is found to be,
\begin{equation}\label{eq.minimum.energy.two.sphere}
E_{m\,\rm min} = E_m(m_{\rm min},\pi) = \frac{m^2}{R^3} \left[ 1 -
\left( \frac{a}{R} \right)^3 \right].
\end{equation}
The relative energy reduction is thus given by the volume ratio of
the two spheres.

Using $E_m^0$ from (\ref{eq.energy.mag.field.rot.sphere}), the energy lowering is now
found to be:
\begin{equation}\label{eq.energy.lowering.of.expulsion}
E_m^0 - E_{m\,\rm min} = \frac{m^2}{R^3}\frac{a^3}{R^3}
=\frac{\vecB_e^2}{4} a^3 = 3\left( \frac{\vecB_e^2}{8\pi}\right)
\left( \frac{4\pi a^3}{3} \right).
\end{equation}
This result is independent of the radius $R$ of the big sphere
introduced to produce the constant external field. It shows that
the energy lowering corresponds to three times the external
magnetic energy in the volume $4\pi a^3/3$ of the perfectly
conducting interior sphere.

\section{The mechanism of flux expulsion}
In 1933 Meissner and Ochsenfeld \cite{meissner}, discussing their experimental discovery, stated that it is understandable that an external magnetic field does not penetrate a superconductor\footnote{Eddy currents induced in accordance with Lenz law do not dissipate because of zero resistivity.} but that the expulsion of a pre-existing field at the phase transition cannot be understood by classical physics. This statement has since been repeated many times. We are not aware, however, of any deeper investigations of what classical electromagnetism predicts regarding the behavior of perfect conductors in this respect, at least not prior to the work of Karlsson \cite{karlsson} and Bad{\'i}a-Maj{\'o}s \cite{badiamajos}. Our theorem strengthens the conclusion of Bad{\'i}a-Maj{\'o}s that a magnetic field is expelled according to classical electromagnetism. Hirsch \cite{hirsch2} has pointed out that the Meissner effect is not explained by BCS theory. Here we briefly speculate on the microscopic physical mechanism of this expulsion.

Assume that a resistive metal sphere is penetrated by a constant
magnetic field. Lower the temperature until the resistance
vanishes. How does the metal sphere expel the magnetic field, or
equivalently, how does it produce surface currents that screen the
external field? According to Forrest \cite{forrest} this can not
be understood from the point of view of classical electrodynamics
since in a perfectly conducting medium the field lines must be
frozen-in. This claim is motivated thus: When the resistivity is
zero there can be no electric field according to Ohm's law, since
this law then predicts infinite current. But if the electric field
is zero the Maxwell equation, $\nabla\times\vecE -
\partial\vecB/\partial\, t =\vecnl$, requires that the time
derivative of the magnetic field is zero. Hence it must be
constant.

This argument is flawed since Ohm's law is not applicable. Inertia, inductive or due to rest mass, prevents infinite acceleration. Instead the system of charged particles undergoes thermal fluctuations and these produce electric and magnetic fields. These fields accelerate charges according to the Lorentz force law. In the
normal situation the corresponding currents and fields remain microscopic. When there is an external magnetic field present the overall energy is lowered if
these microscopic currents correlate and grow to exclude the
external field. According to standard statistical mechanics the
system will then eventually relax to the energy minimum state
consistent with constraints. We note that Alfv\'en and
F\"althammar \cite{BKalfven} state that "in low density plasmas
the concept of frozen-in lines of force is questionable".

Another argument by Forrest \cite{forrest} is that the magnetic
flux through a perfectly conducting current loop is conserved.
Since one can imagine arbitrary current loops in the metal that
just lost its resistivity with a magnetic field inside, the field
must remain fixed, it seems. Is is indeed correct that the flux
through an ideal current loop is conserved, but the actual physical
current loop will not remain intact unless constrained by
non-electromagnetic forces. There will be forces on a loop of current that encloses a
magnetic flux that expands it \cite{BKlandau8}. This is the well know mechanism behind the rail gun, see \eg\ Ess\'en \cite{essen09}. All the little current loops
in the metal will thus expand until they come to the surface where
the expansion stops. In this way the interior field is thinned out
and current concentrates near the surface. So, the flux is
conserved through the loops, but the loops expand.

It was noted early in the history of superconductivity that the Meissner flux expulsion is necessary if superconductors obey normal thermodynamics. Gorter and Casimir \cite{gorter&casimir} observed that the final thermal equilibrium state of a superconductor must be independent of whether the external magnetic field existed inside the body prior to the phase transition or if it was added after the transition already had occurred. Our theorem indicates that it is perfectly natural that an interior magnetic field is expelled in the approach to thermodynamic equilibrium.

\section{Conclusions}
It is amazing that the theorem derived here has not been stated before. Zero resistivity conductors have been known since 1911 and zero resistivity is also considered a good approximation in many plasmas, so the ideal, or perfect conductor, is a well known concept and its magnetic energy minimum ought to be of great importance. It is clear from results above, and further motivated by the examples in the Appendix below, that type I superconductors below their critical field obey the theorem, and the reason that these only have surface current and zero interior field is thus simply minimization of magnetic energy. Naturally there is some other energy involved that is responsible for the zero resistivity itself but apart from being implicitly assumed constant in our variations it is irrelevant to the current investigation.

\appendix
\section{Appendix}
In this Appendix we present further evidence in support of our theorem. The purpose of the calculations presented here is to illuminate and elucidate the physical meaning and the mechanisms behind the expulsion of current and magnetic field from the interior of ideal conductors.

\subsection{Magnetic energy minimization in simple one degree of
freedom model systems} We investigate two simple one degree of freedom
model systems and use them to illustrate how the minimum magnetic
energy theorem works. We take systems in which the magnetic energy $E_m$ can
be calculated exactly so that energy minimization amounts to
minimizing a function of a single variable. The systems are both
related to a system used by Brito and Fiolhais \cite{brito} to
study electric energy.

\subsubsection{Magnetic energy of coaxial cable}
\begin{figure}[ht]
\centering
\includegraphics[width=220pt]{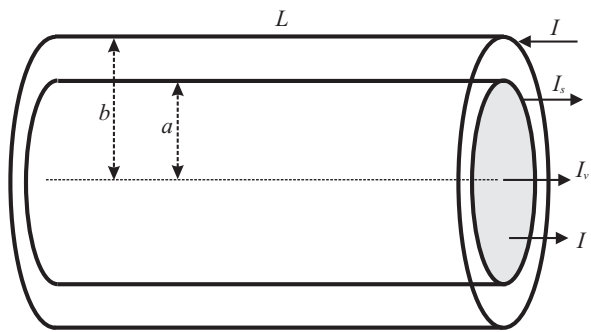}
\caption{\small Notation for the coaxial cable. Magnetic energy is minimized when the current on the inner conductor is pure surface current $I_s=I$, and $I_v=0$.
\label{fig.CoaxCable}}
\end{figure}

The minimum magnetic energy theorem can be illustrated in such a
simple system as a coaxial cable. The cable can be modeled by an
outer cylindrical conducting shell with radius $b$, carrying an
electric current $I$, and a concentric solid cylindrical conductor
with radius $a < b$, carrying the same electric current in the
opposite direction. We now assume that the total current on the inner cylinder is the sum of surface current $I_s$ and bulk interior current $I_v$. See Fig.\ \ref{fig.CoaxCable}

Here we use cylindrical coordinates, $\rho, \varphi, z$
and put $I_s = (1-\eta)I$ at $\rho = a$ for the surface current, and  $I_v=\eta
I$ for the bulk current in $0\le\rho<a$. Then $I=I_s+I_v$, and we get the
magnetic field,
\begin{equation}
\vecB(\rho,\eta) = \frac{2 I}{c}\cdot \left\{
\begin{array}{ll} \displaystyle
\frac{ \eta  \rho}{ a^2}\, \vecephi&\;\; 0 \le \rho <a \\
    & \\ \displaystyle
\frac{1}{\rho}\, \vecephi&\;\; a \le  \rho \le b \\
 & \\ \displaystyle
0 &\;\; b < \rho \\
\end{array} \right.
\label{surface3}
\end{equation}
using Amp\`ere's law. Thus, the magnetic field energy for a length
$L$ of the cable is given by:
\begin{eqnarray}
E_{\rm m}(\eta) & = & \frac{1}{8\pi}\int_{V} \vecB^2 \dfd V \nonumber\\
 & = & \frac{1}{8\pi} \left(\frac{2 I}{c}\right)^2  \left[ \int_{0}^a \left ( \frac{ \eta \rho}{ a^2} \right )^2
  L 2 \pi \rho \, \dfd  \rho +
 \int_{a}^b \left ( \frac{ 1}{\rho} \right )^2 L
2 \pi \rho \, \dfd  \rho \right] \nonumber \\
 & = & \frac{L I^2}{c^2}\left[ \frac{\eta^2}{4} + \ln\!\left(\frac{b}{a}\right)
 \right].
\end{eqnarray}
This magnetic energy reaches its minimum for zero bulk current,
$\eta=0$, corresponding to surface current only, and zero field for
$0 \le \rho <a$.

\subsubsection{Current in sphere due to rigidly rotating charge}
Consider an ideally conducting sphere of radius $R$. Assume that
there is a circulating current in the sphere which can be seen as
the rigid rotation of a charge $Q$ evenly distributed in the thick
spherical shell between $r=a <R$ and $r=R$. The charge density,
\begin{equation}\label{eq.charge.dens.thich.sphere.shell}
\varrho(\vecr) = \left\{ \begin{array}{ll}\displaystyle 0 & \mbox{for $0 \le r < a$} \\
   &  \mbox{  } \\
\displaystyle \frac{3Q}{4\pi(R^3-a^3)}
& \mbox{for $a\le r \le R$} \\
   &  \mbox{  } \\
   0 & \mbox{for $R < r$}
\end{array} \right.
\end{equation}
is assumed to rotate with angular velocity $\vecomega
=\omega\,\vecez$ relative to a an identical charge density of
opposite sign at rest. The current density is then,
\begin{equation}\label{eq.curr.dens.rot}
\vecj(\vecr)=\varrho(\vecr)\,\vecomega\times\vecr,
\end{equation}
and the current, $ I = \frac{Q}{2\pi} \omega , $ passes through a
half plane with the $z$- axis as edge.

\begin{figure}[ht]
\centering
\includegraphics[width=150pt]{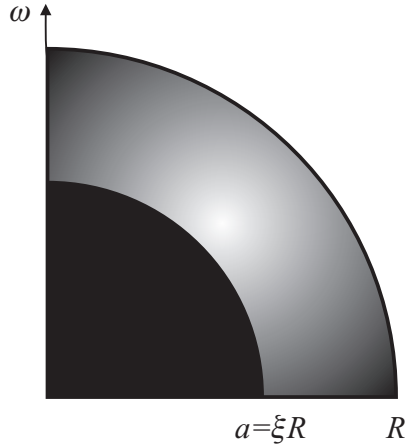}
\caption{\small Some notation for the system considered here. Current density flows in a thick spherical shell as a rigid rotation of constant charge density between $r=a=\xi R$ and $r=R$.
\label{fig.RotCurrSphere}}
\end{figure}

The vector potential produced by this current density can be found
using the methods of Ess\'en \cite{essen89}, see also
\cite{essen96,essen06,essen05}. If we introduce $\xi = a/R$ we
find,
\begin{equation}\label{eq.vec.pot.rot.charge.dens.thich.sphere.shell}
\vecA(\vecr) = \frac{Q}{c} (\vecomega\times\vecr) \cdot
\left\{ \begin{array}{ll}\displaystyle  \frac{(1-\xi^2)}{2(1-\xi^3)R} & \mbox{for $0 \le r < a$} \\
   &  \mbox{  } \\
\displaystyle \frac{R^2}{10 (1-\xi^3)} \left(\frac{5}{R^3} -
3\frac{r^2}{R^5} -2\frac{\xi^5}{r^3} \right)
& \mbox{for $a\le r \le R$} \\
   &  \mbox{  } \\  \displaystyle
 \frac{(1-\xi^5)}{5(1-\xi^3)}\frac{R^2}{r^3} & \mbox{for $R < r$}
\end{array} \right.
\end{equation}
The parameter $\xi=a/R$ is zero, $\xi=0$, for a homogeneous ball
of rotating charge, while $\xi=1$ corresponds to a rotating shell
of surface charge, see Fig. \ref{fig.RotCurrSphere}. Comparing with the vector potential for a
constant field, $\vecA = \half \vecB_0\times\vecr$ we see that,
\begin{equation}\label{eq.middle.field}
\vecB_0 =  \frac{Q\vecomega}{c R} \frac{(1-\xi^2)}{(1-\xi^3)} =
\frac{Q\vecomega}{c R} \frac{(1+\xi)}{(1+\xi+\xi^2)}
\end{equation}
is the field in the central current free region $0\le r<a$.

\subsubsection{Magnetic energy of rotating spherical shell current}
We now calculate the magnetic energy of this system using the
formula,
\begin{equation}\label{eq.energy.intergral.gen}
E_m = \frac{1}{2c} \int \vecj \cdot \vecA \; \dfd V .
\end{equation}
Performing the integration using spherical coordinates gives,
\begin{equation}\label{eq.energy.thick.rot.shell}
E_m(\omega,\xi) = \left(\frac{R\omega}{c}\right)^2 \frac{Q^2}{R}\; f(\xi) ,
\end{equation}
where,
\begin{equation}\label{eq.f.of.xi}
f(\xi) =
\frac{2+4\xi+6\xi^2+8\xi^3+10\xi^4+5\xi^5}{35(1+\xi+\xi^2)^2},
\end{equation}
is a function of the dimensionless parameter $\xi$. Note that $f(0)=2/35$, that $f(1)=1/9$, and that $f(\xi)$ is monotonically increasing, by a factor of almost $2$ in the interval $0$ to $1$.

\begin{figure}[ht]
\centering
\includegraphics[width=200pt]{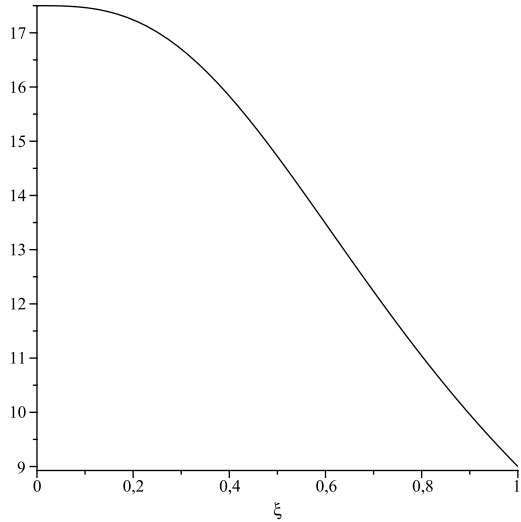}
\caption{\small A graph of the function $1/f(\xi)$ which is
proportional to the Hamiltonian form of the magnetic energy
(\ref{eq.ham.energy.of.model}) of our model system. Note that
$\xi=0$ corresponds to volume (bulk) current and $\xi=1$ to pure
surface current. \label{fofxicurve}}
\end{figure}

This expression for the energy is the Lagrangian form of a kinetic
energy which depends on the generalized velocity
$\omega=\dot\varphi$,
\begin{equation}\label{eq.lagr.energy.thick.rot.shell}
L_m(\dot\varphi,\xi) = \frac{R^2}{c^2} \frac{Q^2}{R}\;
f(\xi)\,\dot\varphi^2 .
\end{equation}
Since the generalized coordinate $\varphi$ does not appear in the
Lagrangian $L_m$ the corresponding generalized momentum (the
angular momentum),
\begin{equation}\label{eq.gen.mom}
p_{\varphi}=\frac{\partial L_m}{\partial \dot\varphi} = 2
\frac{R^2}{c^2} \frac{Q^2}{R}\; f(\xi)\,\dot\varphi ,
\end{equation}
is a conserved quantity. The corresponding Hamiltonian, and relevant, expression for
the magnetic energy is then $H_m = p_{\varphi} \dot\varphi - L_m$,
expressed in terms of $p_{\varphi}$,
\begin{equation}\label{eq.ham.energy.of.model}
E_m(\xi) = H_m(p_{\varphi},\xi) = \frac{c^2}{4} \frac{p_{\varphi}^2}{ Q^2 f(\xi)
R}.
\end{equation}
The function $1/f(\xi)$ is plotted in Fig.\ \ref{fofxicurve}.
We now consider the two energy expressions
(\ref{eq.lagr.energy.thick.rot.shell}) and
(\ref{eq.ham.energy.of.model}) separately.

{\bf Case of constant current:} We first consider the case that
the {\it current,} $I = Q\omega/2\pi$, {\it is constant}. Changing
$\xi$ then means changing the conductor geometry while keeping a
constant total current, or, equivalently, angular velocity
$\omega=\dot\varphi$. One might regard the total current as
flowing in a continuum of circular wires. Changing $\xi$ from zero
to one means changing the distribution of these circular wires
from a bulk distribution in the sphere to a pure surface
distribution, while maintaining constant current. According to a
result by Greiner \cite{BKgreinerCED} a system will tend to {\it
maximize} its magnetic energy when the conductor geometry changes
while currents are kept constant. This has also been discussed in
Ess\'en \cite{essen09}. In conclusion, if currents are kept
constant the magnetic energy
(\ref{eq.lagr.energy.thick.rot.shell}) will tend
(thermodynamically) to a stable equilibrium with at a {\it
maximum} value and we note that this corresponds to a pure surface
current $\xi=1$.

{\bf Case of constant angular momentum:} Assume now that we  pass
to the Hamiltonian (canonical) formalism. Thermodynamically this
type of system should tend to minimize  its phase space energy
(\ref{eq.ham.energy.of.model}) in accordance with ordinary
Maxwell-Boltzmann statistical mechanics. As a function of $\xi$
this Hamiltonian form of the energy $E_m(\xi)$ clearly has a minimum at $\xi
=1$, see Fig.\ \ref{fofxicurve}, corresponding to pure surface
current. In this case therefore there will be current density only
on the {\it surface} in the energy minimizing state. This is in
accordance with the our minimum magnetic energy theorem. It is
notable that {\it both} the assumption of constant current and the
assumption of constant angular momentum lead to a pure {\it
surface current density} as the stable equilibrium.

\subsection{Explicit solutions with minimum magnetic energy}
To further illustrate our theorem we present here three explicit solutions for current
distributions and magnetic fields that minimize the magnetic energy. We do not repeat the solution for a torus since it is a bit lengthy and has been published several times already, probably first by Fock \cite{fock2}, but, independently, several times since then, see \eg\
\cite{launay,carter&loh&po,bhadra,haas1,belevitch&boersma,ivaska&al}. Karlsson \cite{karlsson},  however, was probably the first to notice that the solution minimizes
magnetic energy for constant flux. Dolecek and de Launay \cite{dolecek&delaunay}
verified experimentally that a type I superconducting torus behaves
exactly as the corresponding classical perfectly diamagnetic system for field strength below the critical field. Here we treat three cases all involving a constant external field. For a cylinder, perpendicular to the field, and for a sphere, analytical solutions are found. Finally, for a cube with a space diagonal parallel to the field, we present a numerical solution.

\subsubsection{Cylinder in external perpendicular magnetic field}
Consider an infinite cylindrical ideal conductor with radius $R$
in a external constant perpendicular magnetic field. To get the
vector potential one must solve the following differential
equation,
\begin{equation}
\nabla \times \vecB = \nabla \times \left ( \nabla \times \vecA \right )= 0 .
\label{curl}
\end{equation}
To solve this one should look for the symmetries of the system. We
assume that the external constant magnetic field points in the
$y$-direction and that the cylinder axis coincides with the
$z$-axis. There will then be no dependence on the $z$-coordinate
so the magnetic field is,
\begin{equation}
\vecB =  \frac{1}{\rho} \frac{\partial
A_z}{\partial \varphi}\,\vecerho
 -  \frac{\partial A_z}{\partial \rho}\,\vecephi +  \frac{1}{\rho}
  \left ( \frac{\partial  }{\partial \rho} (\rho A_{\varphi}) -
  \frac{\partial A_{\rho}}{\partial \varphi} \right )\vecez .
\label{magnetic_2}
\end{equation}
Moreover, due to the symmetry of the system, the $z$-component of
the  magnetic field must be zero,
\begin{equation}
\frac{\partial  }{\partial \rho} (\rho A_{\varphi}) -
\frac{\partial A_{\rho}}{\partial \varphi} = 0 .
\label{equacao}
\end{equation}
These assumptions and constraints transform eq.\ (\ref{curl})
into,
\begin{equation}
\nabla \times \vecB = -  \left ( \frac{1}{\rho^2}
\frac{\partial^2 A_z}{\partial \varphi^2} +  \frac{\partial^2
A_z}{\partial \rho^2} +  \frac{1}{\rho}  \frac{\partial
A_z}{\partial \rho}   \right ) \vecez= 0
\end{equation}
which simply is Laplace equation in cylindrical coordinates
($\rho,\varphi,z$).

Before writing down the general  solution,
let us consider the boundary conditions. As $\rho \rightarrow
\infty$, the magnetic field must approach the external one:
$\vecB_0 = B_0 \vecey = B_0 (\vecephi \cos \varphi
+\vecerho \sin \varphi ) $. Furthermore, since the magnetic
field is zero inside the perfect conductor, one concludes from
eq.\ (\ref{magnetic_2}) that the vector potential vector must be
constant inside the cylinder. Therefore, the solution is,
\begin{equation}
A_z = \textrm{const.} + B_0 \left (  \frac{R^2}{\rho} - \rho \right ) \cos \varphi ,
\end{equation}
for $\rho > R$. The magnetic field
outside the cylinder becomes,
\begin{equation}
\vecB = \vecerho\, B_0 \left ( 1 - \frac{R^2}{\rho^2}
\right)\sin \varphi
 + \vecephi\, B_0 \left( 1 + \frac{R^2}{\rho^2} \right )\cos \varphi ,
\label{magnetic}
\end{equation}
which implies that,
\begin{equation}
\vecB(\rho=R) = 2 B_0 \cos \varphi \,\, \vecephi
\label{magnetic8}
\end{equation}
The magnetic field on the cylinder's surface
determines the surface current according to eq.\ (\ref{eq.surf.curr.mag.field}), so we get,
\begin{equation}
\veck =  \frac{c }{2 \pi} B_0  \cos \varphi \,\, \vecez.
\label{surfacecurrent}
\end{equation}
The total current obtained through integration of the surface
current is zero as expected, otherwise the energy would diverge. A
more detailed analysis on this problem can be found in
\cite{zhilichev}.

\subsubsection{Superconducting sphere in constant magnetic field}
Similar calculations can be performed for a superconducting sphere
with radius $R$ in a constant external magnetic field pointing in
the direction of the $z$-axis. As for the cylinder case, eq.\ (\ref{curl}) is considerately simplified using the symmetries of the system. Since the
external constant magnetic field points along the $z$-axis, there won't be any dependence on the $\varphi$ coordinate and the magnetic field along this coordinate must be
zero. Therefore, the magnetic field simplifies to,
\begin{equation}
\vecB =\frac{1}{r \sin \theta} \frac{\partial} {\partial \theta}
 \left( A_\varphi \sin \theta \right) \vecer -  \frac{1}{r}
  \frac{\partial }{\partial r}  \left( A_\varphi r \right) \vecetheta ,
\label{magnetic_3}
\end{equation}
where we use spherical coordinates $r, \theta, \varphi$. Again,
using  these assumptions and constraints eq.\ (\ref{curl})
becomes,
\begin{equation}
\nabla \times \vecB = -  \frac{1}{r} \left [
\frac{\partial }{\partial r} \left ( \frac{\partial }{\partial r}
 \left ( rA_\varphi \right ) \right ) + \frac{1}{r} \frac{\partial }{\partial \theta}
 \left ( \frac{1}{\sin \theta} \frac{\partial }{\partial \theta}
 \left ( A_\varphi \sin \theta \right ) \right ) \right ] \vecephi = 0
\label{curl_2}
\end{equation}
Since the magnetic field is zero inside the sphere, eq.\
(\ref{magnetic_3}) implies that the vector potential has the form,
\begin{equation}
A_\varphi (r < R) = \frac{C}{r \sin \theta} ,
\end{equation}
where $C$ is a constant. To prevent the vector
potential from diverging at $r=0$ and $\theta = 0$, the constant $C$
must be zero. Furthermore, as $r \rightarrow \infty$, the magnetic
field must go to the external field, $\vecB_0 = B_0
\vecez = B_0 (\vecer \cos \theta -\vecetheta \sin \theta ) $.
Therefore, the solution of  eq.\ (\ref{curl_2}) for this case is,
\begin{equation}
A_\varphi (r > R) = \frac{B_0}{2} \left (  r - \frac{R^3}{r^2}
\right ) \sin \theta ,
\end{equation}
which leads to the following magnetic field outside the sphere,
\begin{equation}
\vecB = \vecer\, B_0  \left ( 1 - \frac{R^3}{r^3}  \right) \cos
\theta  - \vecetheta\, B_0 \left( 1 + \frac{1}{2}
\frac{R^3}{r^3} \right )\sin \theta . \label{magnetic_4}
\end{equation}
The magnetic field at the sphere surface
is thus,
\begin{equation}
\vecB = - \frac{3}{2} B_0  \sin \theta  \,\, \vecetheta.
\label{magnetic_5}
\end{equation}
One notes that this is the same field as that of section
\ref{sec.two.sphere.syst} at the surface of the inner sphere when
energy is minimized.

Using eq.\ (\ref{eq.surf.curr.mag.field}), the surface current density becomes,
\begin{equation}
\veck = - \frac{3c}{8\pi} B_0 \sin \theta  \,
\vecephi . \label{current_3}
\end{equation}
Unlike the infinite cylinder in a perpendicular external  field,
the sphere must have a total non-zero electric current, $I =
\frac{3c}{4\pi}R B_0 $, to keep the magnetic field from entering.
A similar approach to this problem can be found in \cite{matute}.

\subsubsection{Perfectly conducting cube in constant magnetic field}
Starting from the magnetic energy functional we have made finite element calculations of the current and magnetic field produced when a perfectly conducting cube is placed in a (previously) constant field. The constant external field is produced by given currents on the surface of a sphere that encloses the cube, as in Subsec.\ \ref{subsec.const.ext.field.sphere}. The cube is placed inside this sphere with one of its space diagonals parallel to the external field.
\begin{figure}
\centering
\includegraphics[width=500pt]{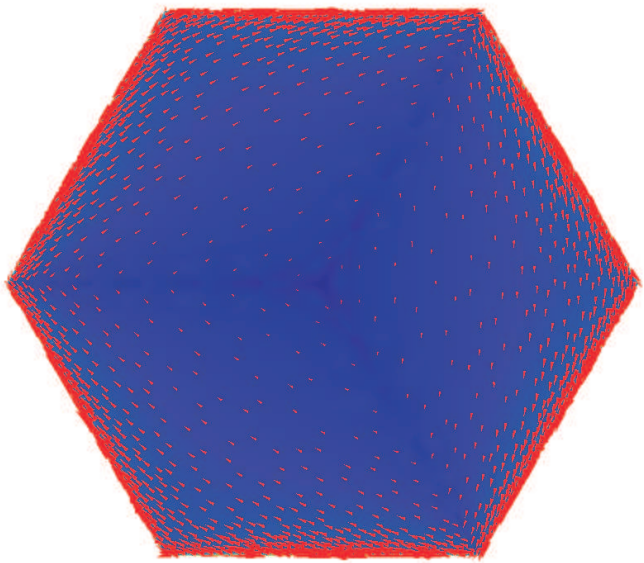}
\caption{\small Result of finite element calculation of the current distribution on an ideally conducting cube in a constant magnetic field. The external field is along a space diagonal of the cube and points up from the figure. The current is on the surface and is indicated by arrowheads. \label{cubesurfcurrentT}}
\end{figure}
\begin{figure}
\centering
\includegraphics[width=500pt]{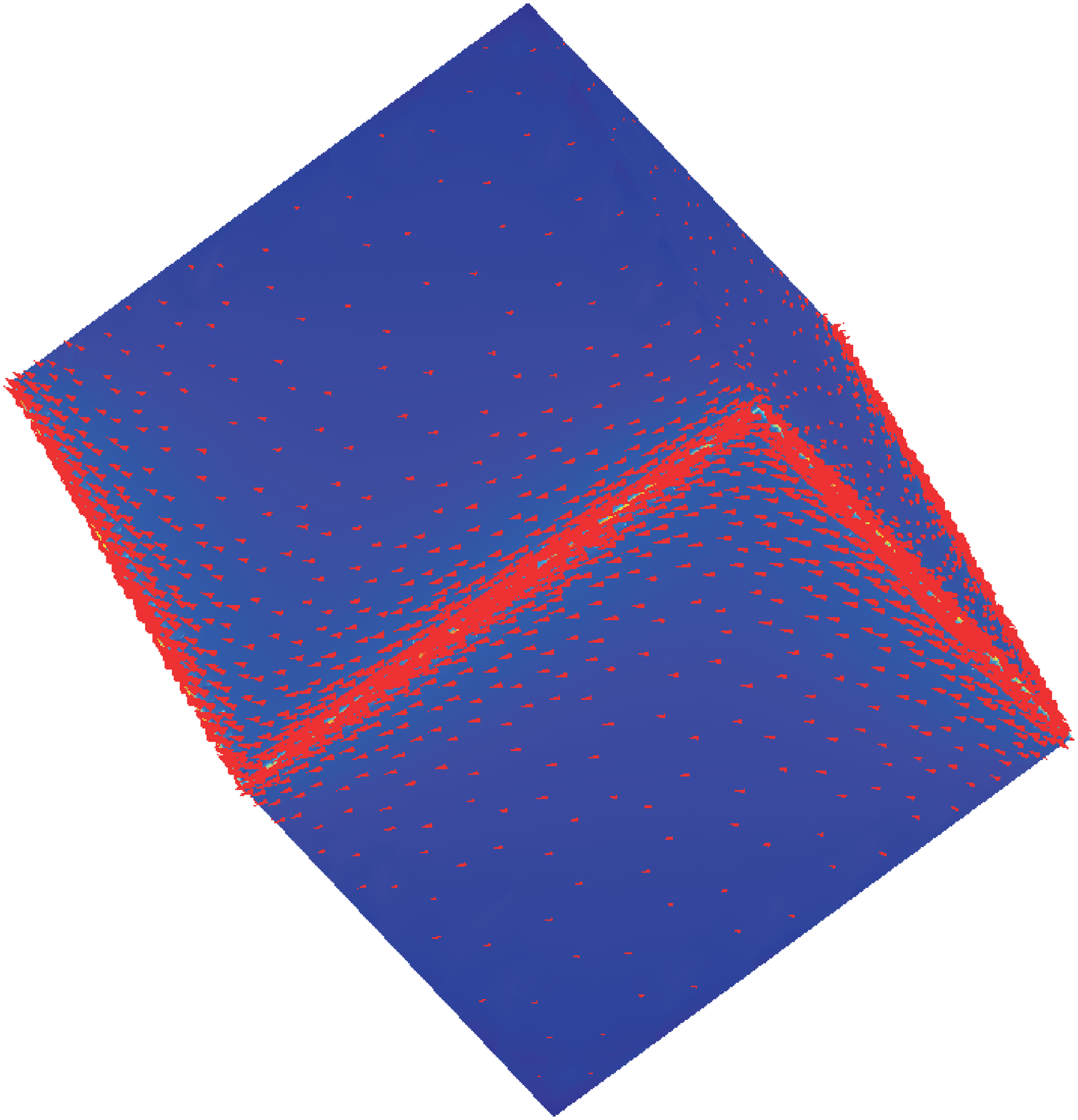}
\caption{\small Same as Figure \ref{cubesurfcurrentT} but with a sideways view of the cube. The current is seen to concentrate along the edges that form a closed path round the cube. The induced surface current expels the external field from the interior of the cube. \label{cubesurfcurrentS}}
\end{figure}
The calculations verify the results of the variational principle: current density and magnetic field become zero inside the cube and the surface current density adjusts to achieve this. In Figures \ref{cubesurfcurrentT} and \ref{cubesurfcurrentS}  we show results for the current distribution of the surface of the cube as seen first from the top (the direction of the external field) and then from the side. One observes that the current density concentrates along the edges that form a closed path round the cube.

We also calculated the magnetic flux through the circular surface enclosed by the equator of the enclosing sphere, which we take to have radius $R=1$ and to produce the constant magnetic field $B_e = 1$ in the interior, when empty. For this case the flux becomes,
\begin{equation}\label{eq.flux.empty.sphere}
\Phi_0 = B_e \pi R^2 = \pi \approx 3.1416
\end{equation}
When an ideally conducting sphere of volume $V=1$, \ie\ radius $a=\sqrt[3]{\frac{3}{4\pi}}$, is placed inside (as in Section \ref{sec.sphere.in.sphere}) the flux is reduced to,
\begin{equation}\label{eq.flux.sphere.sphere}
\Phi_{\rm sp} = \pi \left(1 - \frac{3}{4\pi} \right) \approx 2.3916
\end{equation}
When the sphere is replaced by a cube of volume $V=1$ and a space diagonal parallel to the external fields we find the magnetic flux,
\begin{equation}\label{eq.flux.cube.sphere}
\Phi_{\rm cu}  \approx 2.2733
\end{equation}
numerically. One notes that such a cube excludes more flux than a sphere of the same volume.


\bibliographystyle{unsrt}

\end{document}